\documentclass[usenatbib,usegraphicx]{mn2e}
\usepackage{graphicx}

\title[The Luminosity-Diameter Relations for Globular Clusters and Dwarf
Spheroidal Galaxies]{The Luminosity-Diameter Relations for Globular
Clusters and Dwarf Spheroidal Galaxies}

\author[Sidney van den Bergh]{Sidney van den Bergh\thanks{E-mail:
sidney.vandenbergh@nrc-cnrc.gc.ca}\\
Dominion Astrophysical Observatory, Herzberg Institute of
Astrophysics, 
 National Research Council, 5071 West Saanich Road, \\
 Victoria, B.C., V9E 2E7, Canada \\}

\begin{document}

\date{Received}

\pagerange{\pageref{firstpage}--\pageref{lastpage}} \pubyear{ }

\maketitle

\label{firstpage}

\begin{abstract} 
It is shown that globular clusters and the dwarf spheroidal companions
of the Galaxy have a different distribution of flattening values and appear to occupy adjacent regions of the $M_{v}$ versus log $R_{h}$
plane that can be separated by what will be referred to as the Shapley
line. Surprisingly, typical dwarf spheroidal companions to the Milky Way
System are fainter than the average Galactic globular cluster.

\end{abstract}


\section{INTRODUCTION}

In 1914 Harlow Shapley started his systematic survey of Galactic
globular clusters using the 60-in telescope on Mt. Wilson. The main
finding of this study \citep{sha18ab} was that the Milky Way System is
embedded in a vast halo of globular clusters which is centered in the
direction of Sagittarius. Two decades later \citet{sha38ab} discovery
the Sculptor and Fornax dwarf spheroidal galaxies on plates obtained
with the Bruce Telescope at the Boyden Station of the Harvard
Observatory. Shapley characterized these newly discovered objects as
star clusters of galactic dimensions. This discovery was the first step
in a lengthy exploration that eventually led to the conclusion that the
Galaxy is embedded in a corona of such dwarf spheroidal galaxies. The
exact relationship between these two classes of Galactic satellites
remains a mystery. As \citet{sha43} wrote: ``Two hazy patches [the
Sculptor and Fornax dwarfs] on a photograph have put us in a fog.'' It
is the purpose of the present {\it letter} to try to make a small
contribution to a deeper understanding of the nature of the globular
cluster halo, and of the corona of dwarf spheroidal galaxies, in which
our Milky Way System is embedded.

\section{DATA ON DWARF SPHEROIDALS}

Figure 1 shows a plot of the distribution of the absolute magnitudes
$M_{v}$ of Galactic globular clusters and of the presently known dwarf
spheroidal companions to the Milky Way System (see Table 1) as a
function of their half-light radii. All data for the globular clusters
were drawn from the recent compilation by \citet{macvan05}. The
information on the brightest nearby dwarf spheroidals was taken from
\citet{vdb00}, and data for the fainter nearby dwarf spheroidals was
drawn from \citet{mar08}. The Sagittarius dwarf galaxy was excluded from
Table 1 because both its present luminosity, and its half-light radius,
have probably been affected by Galactic tides.  It should be stressed that the determinations of the ellipticities of the faintest dwarf spheroidal galaxies listed in Table 1 may suffer from significant stochastic noise. In Figure 1 the Galactic globular clusters are plotted as filled red dots, whereas the dwarf
spheroidal companions to the Galaxy are shown as filled blue squares.
To guide the eye the globular clusters and the dwarf spheroidal galaxies in Figure 1 have been separated by the the line\\

$M_{v}$ = 16.2 -14.26 log $R_{h}.$  \hspace*{3.5cm}  (1)\\

\flushleft This relation will subsequently be referred to
as the Shapley line. The faintest dwarf spheroidals plotted in Figure 1 have such a small stellar population that they do not contain a single red giant star.  Some of these objects might actually have fallen slightly above (or the left) of the Shapley line if they had, per chance, harbored a single red giant star.  Nevertheless, with all of the presently available data, Eqn. (1)  provides a slightly more complete way of describing the data than does the \citet{bel07} statement ``that there is a paucity of objects with half-light radii between $\sim$40 and $\sim$100 pc.''  Furthermore, it should be emphasized that the assignment of some of the faintest objects to the dwarf spheroidal class remains provisional until radial velocity information becomes available for a significant number of system memebers.

Information on the globular clusters, which are
located above and to the left of the Shapley line is believed to be
almost complete. On the other hand the data for dwarf spheroidal
galaxies become ever more incomplete as one moves towards the lower
right hand corner of Figure 1. It is expected that many large, faint
(and hence low surface brightness) dwarf spheroidals remain to be
discovered. It is noted in passing that the four extended luminous
clusters that have recently been discovered in the halo of M31 have
$M_{v}$ and $R_{h}$ values \citep{mac06} that place all of these objects
to the left of the Shapley line, i.e. in the globular cluster domain. The recently discovered old extended object M33-EC1 also falls to the left of the Shapley line, i.e. in the globular cluster domain.

\section{DISCUSSION}

Shapley (1938b) wrote: ``If intermediate forms connecting them [i.e.
Sculptor and Fornax] with one of these standard types were found, a
correct interpretation would be facilitated.'' The data plotted in
Figure 1 suggest that, at least in the Milky Way System, such
indetmediate-type objects are lacking. The most striking difference
between globular clusters and dwarf spheroidal galaxies is that the
diameters of the latter are typically one or two orders of magnitude
larger than those of the former. A second difference revealed by Figure
1 is that the known dwarf spheroidals are spread out over a range of
$10^{5}$ in luminosity, whereas the luminosity distribution of globular
clusters is strongly peaked at $M_{v}$ $\simeq$ -7.5. A Kolmogorov-
Smirnov test shows that there is only an 8\% probability that the
observed luminosities of globular clusters and of dwarf spheroidals were
drawn from the same parent population. The existing sample of dwarf
spheroidal companions to the Galaxy is almost certainly incomplete in
the lower right hand corner of Figure 1. Future discoveries are
therefore expected to reduce the probability that dwarf spheroidals and
globular clusters were drawn from the same parent population. It is of
interest to note that the median luminosity of the known dwarf
spheroidal galaxies in Table 1 is $M_{v}^{*}$ $\sim$ -6.5, whereas the
median luminosity of Galactic globular clusters is $M_{v}^{*}$ $\sim$
-7.5. In other words typical dwarf spheroidal {\it galaxies} are fainter
than the average globular {\it cluster}. This difference is likely to
increase as more very faint and low surface brightness dwarf spheroidal
companions to the Galaxy are discovered. At lease part of this
difference is, no doubt, due to the fact that many low-mass globular
clusters were destroyed by stellar-dynamical evaporation
\citep{mclfal08}. Since the mass loss rate -dM/dt $\propto$ $R_{h}^
{-3/2}$, such mass loss will affect compact clusters much more than
extended dwarf spheroidals. In other words, one cannot yet exclude the
possibility that dwarf spheroidals and globular clusters might initially
have formed with similar luminosity distributions. The conclusion that
typical dwarf spheroidal galaxies are quite faint, and therefore
difficult to observe, fits very comfortably into the framework of a
hierarchical clustering scenario in which massive galaxies, such as the
Milky Way System, should be surrounded by large numbers of satellite
dark matter dominated halos \citet{kauwhigui93}, \citet{kly99},
\citet{moo99}.

Data on the normalized distribution of flattening values for the 100 globular clusters \citep{vdb08} that lie above and to the left of the Shapley line, and for the 21 dwarf spheroidals \citep{mar08} that are situated below and to the right of the Shapley line are listed in Table 2 and plotted in Figure 2.  This figure shows that the objects below the Shapley line are significantly more flattened than are those that lie above it.  A Kolmogorov-Smirnov test shows that there is a $<0.01 \%$ probability that these globular clusters and swarf spheroidal galaxies were drawn from the same parent population of flattening values.  The reason(s) for this difference are not yet fully understood.  Rotation would suggest implausibly small internal velocity distributions and tidal deformation would require very eccentric orbits for some dwarf spheroidal companions to the Galaxy. Alternatively the high observed flattening of the dwarf spheroidal galaxies might perhaps be due to the fact that they are
embedded in tri-axial dark matter mini halos. The existence of such dark
matter halos might also contribute to the fact that the stars in dwarf
spheroidals appear to exhibit both a greater age spread, and a larger
range in metallicities, than do those in most globular clusters.

\section{CONCLUSIONS}

The Milky Way system is embedded in a halo of globular clusters and in a more extended corona of dwarf spheroidals galaxies.  These two classes of objects show a highly significant difference in their average flattenings, with dwarf spheroidals being more elongated than globular clusters.  Furthermore dwarf spheroidal companions to the galaxy and galactic globular clusters are located in distinct regions of the $M_{v}$ versus log $R_{h}$ diagram.   Available data allow one to draw a line (the Shapley line) that separates Galactic globular clusters from presently known dwarf spheroidal galaxies. The extended luminous globular clusters in M31 and M33 fall inside the same domain as do Galactic globular clusters. It is also pointed out that the median luminosity of the dwarf spheroidal companions to the Galaxy is lower than that of globular clusters. Still undiscovered dwarf spheroidal companions to the Galaxy are expected to be both larger and fainter than those which are already known. New
discoveries are therefore expected to widen the gap between the median
luminosities of globular {\it clusters} and dwarf spheroidal {\it
galaxies}. In other words dwarf spheroidal {\it galaxies} are, on average,
significantly fainter than typical globular star {\it clusters}.

I am indebted to Giuseppe Bertin for a discussion of the properties of
globular clusters and to Bonnie Bullock, Brenda Parrish and Jason
Shrivell for technical assistance. Also, I would like to thank a particularly helpful referee who emphasized how the positions of the faintest objects in Figure 1 might be affected by the presence of a single red giant star.

\begin{figure}
\begin{center}
\includegraphics[angle=-90,width=8.0cm]{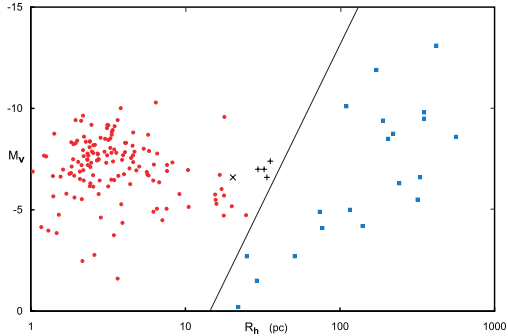}
\end{center}
\caption[]{The figure shows a clear-cut separation between the
distribution of Galactic globular clusters (filled red circles) and
dwarf spheroidal companions to the Galaxy (filled blue squares). A
line separating these two types of objects is given by Eqn. (1). The
data are most incomplete in the lower right-hand corner of the diagram,
i.e. for the faintest and largest objects. Note the striking difference
between the luminosity distributions of globular clusters and dwarf
spheroidal galaxies. Four luminous extended globular clusters in the
outskirts of M31 \citep{mac06} are shown as plus signs.  The extended
cluster M33-EC1 is plotted as a cross.}
\end{figure}

\begin{figure}
\begin{center}
\includegraphics[width=8.0cm]{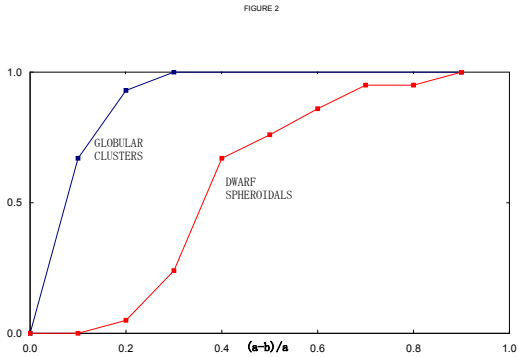}
\end{center}
\caption[]{Normalized frequency distribution of flattening values for objects situated above and to the left of the Shapley line (globular clusters) and for those situated to the right and below this line (dwarf spheroidal galaxies).  The figure shows that dwarf spheroidals galaxies are, on average, much more flattened than are globular clusters.}
\end{figure}

\begin{table}
\caption[]{Data for nearby dwarf spheroidals}
\begin{tabular}[]{lrrr}
\hline
Name & $\varepsilon$ &  $M_{v}$ & $R_{h}$(pc)\\
\hline

Boo I     &  0.39  &  -6.3  & 242\\
Boo II    &  0.21  &  -2.7  &  51\\
CVn I     &  0.39  &  -8.6  &  564\\
CVn II    &  0.52  & -4.9   &  74\\
Car       &  0.33  &  -9.4  &  190\\
Com       &  0.38  &  -4.1  &  77\\
Dra       &  0.31  &  -8.75 &  221\\
For       &  0.30  & -13.1  &  420\\
Her       &  0.68  &  -6.6  &  330\\
Leo I     &  0.21  & -11.9  &  172\\
Leo II    &  0.13  & -10.1  &  110\\
Leo IV    &  0.22  &  -5.0  &  116\\
Leo T     &  0.29  &  ...   &  178\\
Scl       &  0.35  &  -9.8  &  350\\
Seg 1     &  0.48  &  -1.5  &  29\\
Sex       &  0.35  &  -9.5  & 350\\
UMa I     &  0.80  &  -5.5  &  318\\
UMa II    &  0.63  &  -4.2  &  140\\
UMi       &  0.50  &  -8.5  &  205\\
Wil 1     &  0.47  &  -2.7  &  25\\
S1058     &  0.38  &  -0.2  &  22\\

\hline
\end{tabular}
\end{table}

\begin{table}
\caption[]{Normalized frequency distribution of flattening distributions for Galactic globular clusters and dwarf spheroidal companions to the Galaxy}
\begin{tabular}[]{rrr}
\hline
(a-b)/a &  globular clusters  &   dwarf spheroidal galaxies\\
\hline

0.00      &        0.00       &          0.00\\
0.10      &        0.67       &          0.00\\
0.20      &        0.93       &          0.05\\
0.30      &        1.00       &          0.24\\
0.40      &        1.00       &          0.67\\
0.50      &        1.00       &          0.76\\
0.60      &        1.00       &          0.86\\
0.70      &        1.00       &          0.95\\
0.80      &        1.00       &          0.95\\
0.90      &        1.00       &          1.00\\

\hline
\end{tabular}
\end{table}

\label{lastpage}

\begin{thebibliography}{}

\bibitem[\protect\citeauthoryear{Belokurov et al.}{2007}]{bel07} Belokurov, V. et al. 2007, ApJ, 654, 897


\bibitem[\protect\citeauthoryear{Kaufmann, White \&
Guideroni}{1993}]{kauwhigui93} Kauffmann, G., White, S. D. M. \&
Guideroni, B. 1993, MRAS. 264, 201

\bibitem[\protect\citeauthoryear{Klypin et al.}{1999}]{kly99} Klypin,
A., Kravtsov, A. V., Valenzuela, O. \& Prada, F. 1999, ApJ, 522, 82

\bibitem[\protect\citeauthoryear{Mackey et al.}{2006}]{mac06} Mackey, A.
D., Huxor, A., Ferguson, A. M. N., Tanvir, N. R., Irwin, M., Ibata, R.,
Bridges, T., Johnson, R. A. \& Lewis, G, 2006, ApJ, 653, L105

\bibitem[\protect\citeauthoryear{Mackey \& van den
Bergh}{2005}]{macvan05} Mackey, A. D. \& van den Bergh, S. 2005, MNRAS,
360, 631 

\bibitem[\protect\citeauthoryear{Martin et al.}{2008}]{mar08} Martin, N.
F., de Jong, J. T. \& Rix, H-W. 2008, ApJ (in press = arXiv:0805.2945v2)

\bibitem[\protect\citeauthoryear{McLaughlin \& Fall}{2008}]{mclfal08}
McLaughlin, D. E. \& Fall, S. M. 2008, ApJ, 679, 1272 

\bibitem[\protect\citeauthoryear{Moore et al.}{1999}]{moo99} Moore, B.,
Ghigna, S., Governato, F.,Lake, G., Quinn, T, Stadel, J. \& Tozzi, P.
1999, ApJ, 524, L19

\bibitem[\protect\citeauthoryear{Shapley}{1918ab}]{sha18ab} Shapley, H.
1918a ApJ 48, 154

\bibitem[\protect\citeauthoryear{Shapley}{1918ab}]{sha18ab} Shapley, H.
1918b, Proc. Nat. Acad. Sci. 4, 224 

\bibitem[\protect\citeauthoryear{Shapley's}{1938ab}]{sha38ab} Shapley,
H. 1938a, Harvard Bull. 908,1 

\bibitem[\protect\citeauthoryear{Shapley's}{1938ab}]{sha38ab} Shapley,
H. 1938b, Nature, 142, 715 

\bibitem[\protect\citeauthoryear{Shapley}{1943}]{sha43} Shapley, H. 1943
Galaxies (Philadelphia: Blakiston),p.142 

Stonkut$\dot{e}$, R. et al. 2008, AJ, 135, 1482

\bibitem[\protect\citeauthoryear{van den Bergh}{2000}]{vdb00} van den
Bergh, S. 2000 The Galaxies of the Local Group (Cambridge: Cambridge
University Press)

\bibitem[\protect\citeauthoryear{van den Bergh}{2008}]{vdb08} van den
Bergh, S. 2008, MNRAS, 385, L20.

\end{thebibliography}
\end{document}